\documentclass{article}
\usepackage[english]{babel}

\usepackage{graphicx}

\usepackage{geometry}
\usepackage{indentfirst}
\usepackage{tabularx}     
\usepackage{amsmath}      
\usepackage{amsfonts}            
\usepackage{mathrsfs}            

\usepackage{color}    
\usepackage{graphicx}    
\usepackage{ulem}    

\usepackage{hyperref}
\hypersetup{colorlinks=true,linkcolor=blue,filecolor=magenta,urlcolor=cyan,}
\usepackage{multirow}

\title{Theory and Experiment of Dynamic Structure 
in Four-strategy Game\footnote{An extended abstract has appeared in the 15th international conference on game theory and management (GTM2021), June 23 - 25, 2021 St. Petersburg, Russia.  WZJ designed the research, ZSJ carried out the experiments and primer data analysis (results see her master thesis, YQM and WYJ contributed on associated experiment , critical data analysis and manuscript writing. We thanks Pan Gang, Li Shijian, Fan Jijian, Jiamei Lian, Zheng Jie, Zhang Jianbo, Ke Rongzhu, Xu Bin, Chen Fadong and Li Jingyuan for helpful comments. WZJ thanks Shan Lixia's for the help on manuscript preparing.}}
 \author{Wang Zhijian$^a$, Zhou Shujie$^a$, Yao Qinmei$^a$, Wang Yijia$^{b}$\\
$^a$Experimental Social Science Laboratory, Zhejiang University, Hangzhou, China \\
$^b$Ant Group, Hangzhou, China} 
\date{\today}

\begin{document}

\maketitle
%
%
%
%
%
 
Game dynamics theory, as any field of science, 
the consistency of theory and experiment is essential. 
In the past 10 years, important progress has been made in the merging of the theory and experiment in this field, in which dynamics cycle is the presentation. However, the merging works have not got rid of the constraints of Euclidean two-dimensional cycle so far. This paper uses a classic four-strategy game  to study the dynamic structure (non-Euclidean superplane cycle). The consistency is in significant between the three ways: (1) the analytical results from evolutionary dynamics equations, (2) agent-based simulation results from learning models and (3) laboratory results from human subjects game experiments.  The consistency suggests that, game dynamic structure could be quantitatively predictable, observable and controllable in general. 
 




\tableofcontents

\section{Introduction}

\subsection{The background}
For a discipline of science, 
the consistency of the theory and experiment is essential, of which the accuracy and the reality are the two fundamental aspects.
As a discipline of science, game theory, which attempts to explain the strategy interactions between  human subjects, is not an exception. 

Game statics theory, which is called also as classical game theory, is the main stream in game theory. In classical game theory, Nash equilibrium is the central concept. This theoretical concept was established near 1950. 
Till 1987, the human subject game experiment by O'Neill \cite{ONeill1987} 
provided the first illustration
that laboratory human strategy behaviour 
can be accurately captured by the concept. 
Since then, this game has been extensively repeated in various setting experiment \cite{Binmore2001Minimax}\cite{Yoshitaka2013Minimax}. 
Surrounding the central concept, now, 
human subject behaviour game theory and experiment has become a fruitful academic branch \cite{Behavioral2003}. Moreover, game statics theory has been widely applied in real live policy design (mechanism design) for desired social and economical aims.

Game dynamics theory, which is based on evolutionary game theory, is less developed during last 50 years. In the past 10 years, important progress has been made in the merging of the theory and experiment  \cite{dan2014,wang2014social,wang2014,dan2021price}. However, the theoretical inferences and experimental measurements have not got rid of the constraints of two-dimensional Euclidean space so far (e.g., the theoretical expectation is 2-d Cason et al. 2010 \cite{dan2010tasp}\cite{wang2014}\cite{dan2014}\cite{wang2014social} or measured in 2-d Cason et al. 2020\cite{dan2021price}). 
There lacks of sufficient evidences which can bridge experiment and theory of high dimensional game dynamics well.  

Recently, it is found that, the eigen mode (invariant manifold) plays a cruel role in human subject game experiments \cite{WY2020}. In the O'Neill game experiments, which state space having 8 dimensions (\cite{ONeill1987}, \cite{Binmore2001Minimax}, \cite{Yoshitaka2013Minimax}), 
the dynamics pattern in the experiments can be accurately interpreted by the eigen mode of game dynamics equations \cite{WY2020}. 
Guided by the eigen system analysis of the replicator dynamics equations for the O'Neill game,
the authors applied the complex eigenvectors structure to interpret the dynamics structures in the long-existed hunam game experiment data. The logic chain, which root in nonlinear dynamics theory (see Chap 6 in \cite{2019Roussel}), is following

\begin{itemize}
\item For a give game, the game dynamics system can be expressed as a velocity vector field \cite{2011Sandholm}\cite{dan2016} as 
\begin{equation}
    \dot{x} = f(x) 
\end{equation}
in which $x\in R^N$ and $N$ is the dimension of the strategy space.  
\item The Nash equilibrium is a singular point of the vector field (a point
where $\dot{x} = 0$). In linear approximation, near the singular point, the dynamics can be expressed as  \cite{2011Sandholm}\cite{dan2016} 
$$\dot{x} = J  x$$
in which $J$ is the Jacobian (character matrix) at the singular point.
\item Suppose that, $\xi_i$ is the eigenvector associated to the eigenvalue $\lambda_i$ of the diagonalizable $J$; and suppose that, an initial condition can be expressed as  
   $x(0) = \sum_{i=1}^N a_i {\xi}_i$
then, the evolution trajectory can be expressed as 
\begin{equation}\label{eq:eigdeco2}
x(t) = \sum_{i=1}^N e^{\lambda_1t} a_i  {\xi}_i.   
\end{equation}
Here, the eigenvector $\xi_i$ describes an eigen mode, which is a normal mode in an oscillating system (which may have many components), being one in which all parts of the components are oscillating with the same frequency $\lambda_i$. 
\item If the system exists an invariant manifold (eigen mode, a persist periodic obit, a persist loop), the manifold could be captured by a complex eigenvectors. For a given complex eigenvector, disregarding the dimension of the game, there exist a measurement (constructed as eigencycle set in theory, as angular momentum in experiment time series) to identify the invariant manifold. Here, the eigenvectors play the cruel role. 
\item Such that, a high dimension game dynamics structure is expected to be theoretically predictable and experimentally measurable. 
\end{itemize}  
This logic chain has test out the dynamics structure in significant in the existed data (\cite{ONeill1987}, \cite{Binmore2001Minimax}, \cite{Yoshitaka2013Minimax}) reported by \cite{WY2020}. 

\subsection{Motivation and the game selection}
On the motivation and the game selection of this study, we consider following points:  
\begin{enumerate}
\item Although in the long-existed O'Neill game experiments \cite{ONeill1987}\cite{Binmore2001Minimax}\cite{Yoshitaka2013Minimax}  
the high dimensional dynamics pattern meets the theory incredibly well \cite{WY2020}, but the evidence is unique\footnote{After the experiment of this study carried out, Yao \cite{2021Qinmei} reports the similar result as those found in the existed O'Neill game experiments reported \cite{WY2020}. In the independent experiments,  which is a one population 5-strategy symmetric game having unique pair of complex eigenvectors (eigen mode), the dynamics structure was test out; The experiment results is consistence the expectation of the eigencycles from evolutionary dynamics theory well in significant \cite{2021Qinmei}.}. Whether the meet is only by coincidence? This is a puzzle.
\item On game dynamics cyclic pattern, existed the theoretical inferences and experimental measurements have not got rid of the constraints of Euclidean two-dimensional space so far (for details of related literature, see Discussion \ref{sec:diss}). 
\item In game dynamics theory \cite{dan2016}\cite{2011Sandholm}, to obtain a superplane cycle, the game state space must have 3 independent variable. The candidate game is one population 4 strategy game, or two population 2 + 3 strategy, or three population 2 + 2 + 2 game. For simplicity and no lost the generality, we choose a symmetric 4 strategy game. 
\end{enumerate}
So in this study, we limit ourselves on a superplane cyclic game to investigate the consistence of the theory and experiments on the dynamics structure. 
  
Hofbauer and Sigmund (1998) \cite{1998Sigmund}\cite{2011Sandholm} has designed 
an 4 strategy superplane cyclic game, which is a symmetric game (one population game), and its
payoff matrix can be expressed as Table \ref{tab:gameRAmatrix}.  
\begin{table} 
\caption{The 4 strategy matrix}
\begin{center}\begin{tabular}{c|cccc}
&$s_1$ &$s_2$ &$s_3$ &$s_4$  \\
\hline
$s_1$ &	 0 & 0 & 0 & $a$\\
$s_2$ &	 	 1 & 0 & 0 & 0   \\
$s_3$ &	 	 0 & 1 & 0 & 0   \\
$s_4$ &	 	 0 & 0 & 1 & 0  \\ 
\end{tabular}\\ 
\end{center}
\label{tab:gameRAmatrix}
\end{table}
We denoted the payoff matrix as $A$. The element of this matrix $A(1,4)$ is $a$. 
To our study aim, we define $a$ as a positive and is a real number. 
We will control $a$
to illustrate whether the dynamics structure controlled by $a$.
 
 
   
In the state space (denoted as $S$), we assign one by one from the $(x_1,x_2,x_3,x_4) \in S$ as the 
strategy probability of $(s_1,s_2,s_3,s_4)$ in the one poplulation. 
Then, at any time ($t$), the social state of the dynamics system must be
a point the 4-dimension space $S(x_1,x_2,x_3,x_4)$.
In this 4-dimension space, it can be verified that, 
the unique mixed strategy Nash equilibrium is at

\begin{equation}\label{eq:nash_equilibrium}
    x^* = (x_1^*,x_2^*,x_3^*,x_4^*)=\frac{1}{3a+1}\big(a,a,a,1\big).
\end{equation}

On the organisation of this report --- In section 2,  we report the results from three ways,  
(1) we deduce the theoretical expectation by eigen system analysis. (2)
we introduce the results from agent-based simulations. (3)
we introduce the results from human subjects game experiments.
Then we test whether the results are consistence. 
In section 4, we summarise the conclusion, the contribution,  the related works and some suggestions on  
the further research.

\section{Results}\label{sec:result}
\subsection{Results from dynamics models}\label{sec:res_theo} 

\subsubsection{The eigen mode of the evolutionary dynamics}
To investigate the dynamic behaviors in laboratory experiment game,
we begin with using the replicator dynamics equations \cite{taylor1978evolutionary}:
\begin{equation}\label{eq:repliequl}
    \Dot{x}_i= x_i (U_i - {\overline{U}}) 
\end{equation}
in which,
$x_i$ is the $i$-th strategy player's probability
in the population where the $i$-th strategy player included, and
$\Dot{x}_i$ is the evolution velocity of the probability;
 $U_i$ the payoff of the $i$-th strategy player,
and $\overline{U}$ is the average payoff of
the full population.
This is a time invariant dynamics system. 
Suppose that, the motion of the strategy vector $x$ is close the equilibrium and the linear approximation of 
dynamical system is validate, we can obtain the eigen system from the Jacobian (character matrix)\cite{2011Sandholm}\cite{dan2016}.  

The Jacobian at the  unique mixed strategy Nash equilibrium of the dynamics in Eq. (\ref{eq:repliequl}) can be calculated, and the result is  
$$  J = \frac{a}{(3a+1)^2} \left(\begin{array}{cccc} -2\,a & -2\,a & -a-1 & 2\,a^2\\ a+1 & -2\,a & -a-1 & -a\,\left(a+1\right)\\ -2\,a & a+1 & -a-1 & -a\,\left(a+1\right)\\ -2 & -2 & 2 & -a-1 \end{array}\right).$$
By the Jacobian, we can calculate the eigenvalues $\lambda$ and
 their related eigenvector $v$'s components
 $(\eta_1,\eta_2,\eta_3,\eta_4)$ explicitly.
The eigenvalues $\lambda$  are  
\begin{equation}\label{eq:lambda_matrix}
    \lambda =\frac{-a}{3\,a+1} \left(\begin{array}{cccc}
    i & 0 & 0 & 0\\ 0 & -i & 0 & 0\\ 0 & 0 & 1 & 0\\ 0 & 0 & 0 & 1 \end{array}\right) 
\end{equation} 
In the narrative of dynamics system theory, this game is neutral, because the maximum of the real part of the eigenvalues is 0. That is to say, in the replicator dynamics hypothesis, there is a pair of
purely imaginary eigenvalues, we will therefore have a two-dimensional center manifold associated with these eigenvalues.  

Having the eigenvalue, we can have their related eigenvectors:  
%
\begin{equation}\label{eq:v}
    v=  \left(\begin{array}{cccc} \frac{1}{4}({a}-{1}+{3ai}+{i}) & \frac{1}{4}({a}-{1}-{3ai}-{i}) & 0 & 1\\
    -\frac{a}{2}-\frac{1}{2} & -\frac{a}{2}-\frac{1}{2} & a &0 \\ \frac{1}{4}({a}-{1}-{3ai}-{i}) & \frac{1}{4}({a}-{1}+{3ai}+{i}) & 0 &1 \\ 1 & 1 & 1 & 0 \end{array}\right)
\end{equation} 
%
It is worth to notice that, there is a pair of conjunction complex eigenvalues, and naturally, their associated eigenvectors are of a pair of conjunction complex eigenvectors. 
These complex eigenvectors determine the dynamics structure of the game. 
With this explicitly expression of the eigenvector shown in Eq. (\ref{eq:v}), the eigencycle can be obtain as following.  
\subsubsection{The eigencycle and rotation axis}
%
%
Follow \cite{WY2020}, for a $N$-dimensional dynamics system, 
a eigencycle is constructed by
two components $(\eta_m,\eta_n)$ within one normalised eigenvector
${v}_i = (\eta_1, ..., \eta_m, ..., \eta_n, ... \eta_N)^T$. The eigencycle, marked as $\sigma^{(mn)}$, and defined as follows:
\begin{equation}\label{eq:the_ecyc}
      \sigma^{(mn)}=\pi \cdot ||\eta_m|| \cdot ||\eta_n|| \cdot  \sin\left(\arg(\eta_m)-\arg(\eta_n)\right)
\end{equation}
in which, the superscript $(mn)$ is the index of the 2-d subspace
where $m$  and $n$ are the abscissa
and the ordinate dimension respectively;
$||\eta_m||$ and $\arg(\eta_m)$ indicates the amplitude
and the phase angle of the $\eta$, respectively.
$\sigma^{(mn)}$ can determine the direction of the eigencycle
and the amplitude of the eigencycle.
According to this formula,  the eigencycle values of the eigenvectors shown in Eq. (\ref{eq:v}) are 
\\ 
%
\begin{equation}\label{eq:sigmamatrix}
    \sigma = \left(\begin{array}{cccc} \frac{a}{2}+\frac{1}{2}&-\frac{a}{2}-\frac{1}{2}&0&0\\ \beta &-\beta&0&0\\ -1&1&0&0\\ \frac{a}{2}+\frac{1}{2}&-\frac{a}{2}-\frac{1}{2}&0&0\\ 0&0&0&0\\ 1&-1&0&0 \end{array}\right) 
\end{equation}
$$  
$$
in which $\beta=\frac{1}{3\,a+1}\Big[{4\,\sin\left({\arg}\left(a\,\left(\frac{1}{4}-\frac{3}{4}{}\mathrm{i}\right)-\frac{1}{4}-\frac{1}{4}{}\mathrm{i}\right)- {\arg}\left(a\,\left(\frac{1}{4}+\frac{3}{4}{}\mathrm{i}\right)-\frac{1}{4}+\frac{1}{4}{}\mathrm{i}\right)\right)\,\left(\frac{5\,a^2}{8}+\frac{a}{4}+\frac{1}{8}\right)}\Big]$. 
It is worth to notice that (1) for the eigencycle values referring to a real eigenvalue are zero, because $\arg(\eta_m)$ equals to $\arg(\eta_n)$.
In words, the related eigenvector components 
have no phase difference. (2) The pair of the eigencycle set, which associated to a pair complex conjunction eigenvectors, their value are oppose. 
In this study case, it is interesting to notice that, 
disregarding the value $a$, following relationship hold
\begin{eqnarray}
    \sigma^{24} &=& 0,  \label{eq:srtict_pm0}\\ 
    \sigma^{14} &=& - \sigma^{34}, \label{eq:srtict_pm1}\\ 
    \sigma^{12} &=& \sigma^{23}.\label{eq:srtict_pm2}  
\end{eqnarray} 
In section \ref{sec:res_human} about the result from human subject experiments, these  relationship (the independent of $a$ relationship, or called as $a$-invariant relationship) will be test statistically (result is shown in Fig.  \ref{fig:strict_pm}). 

\begin{figure} 
\centering
 
\includegraphics [width=4.5in]{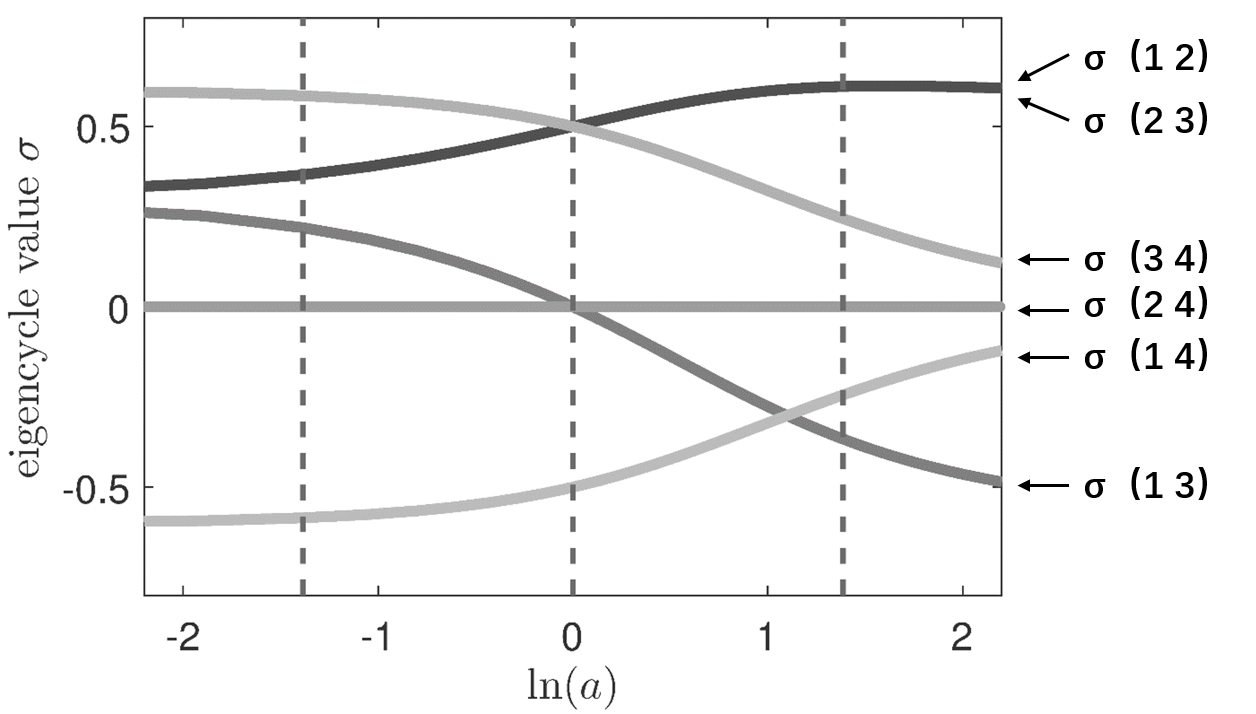}\\  
 
\caption{Eigencycle value referring to $a$. In order to the symmetric visibility, the holizon axis is scaled with the natural log function. These curves represent the eigencycles values of the replicator dynamics shown in Eq.~(\ref{eq:repliequl}) for the 4 strategy (symmetric one population) game with the payoff matrix shown in Table \ref{tab:gameRAmatrix}. The left most and the right most dash line indicate the $a=[1/4]$ and $a=[4]$ condition (treatment), which will be investigated in theory and experiments in this study, respectively. }
\label{fig:ec_a}
\end{figure}

\begin{figure} 
\centering
 
\includegraphics[scale=0.55]{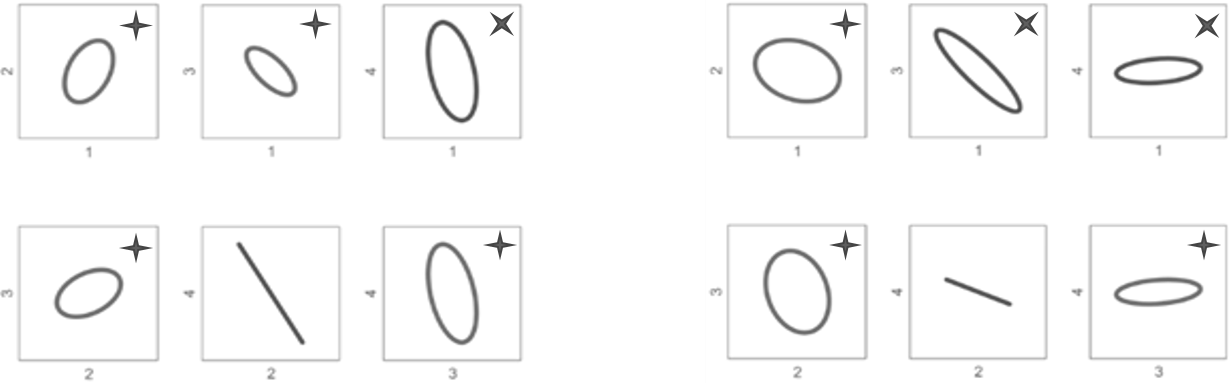} 
 
\caption{The geometric presentation of the eigencycle values $\sigma_{mn}$ of the six  2-dimensional subspace. Results come from the replicator dynamics (see Eq. (\ref{eq:repliequl})) for the game (see Table \ref{tab:gameRAmatrix}), in which Left panel $a=[1/4]$ and Right panel $a=[4]$. The symbol $\times$ (blue in electronic version) indicates the cycle is clockwise being negative value, and the symbol $+$ (red in electronic version) indicates the cycle is counter clockwise being positive value. The relative value of areas of the six cycles is the relative value of the six eigencycle. The angular momentum $L_{mn}$ measured the six subspace in time series of experiment should be proportional to these  $\sigma_{mn}$ value (the proof has been shown in \cite{WY2020}).}
\label{fig:lissa}
\end{figure}

Referring to \cite{WY2020}, we interpretation the eigencycle and the parameter selection for this study as following points. 
\begin{itemize}
\item \textbf{Number of eigencycle:}
      The number of the independent eigencycle in this study is 6. Referring to its definition, for an $N$-dimensional system, there are $N(N-1)/2$ independent eigencycles
      corresponding to a given $N$-component normalised eigenvector.
      Because there are a total of $N^2$  pairwise
      combination of each component in an $N$-dimensional eigenvector.
      Considering the $N$  self-combination of
      $(\eta_m,\eta_m)$ are trivial ($\sigma^{(mm)}$ = 0),
      and $(\eta_n,\eta_m)$ and $(\eta_m,\eta_n)$
      is just a simple reverse ($\sigma^{(mn)}$ = $-\sigma^{(nm)}$),
      so that only $N(N\!-\!1)/2$ combinations are remained. In this study, $N=4$, so the number of the eigencycle is 6.
\item \textbf{Eigencycle set:} In this study case, 
        their is only one independent eigencycle set (see the first column in the eigenvector matrix $\sigma$). Because, 
        there has only one pair conjunction complex eigenvalues (see the 1st and the 2nd  diagonal element  of the eigenvalue matrix in $\lambda$). 
        The associated complex eigenvector 
        is a pair of conjunction complex vector (see 1st and 2nd column of the eigenvector matrix in $v$).
        Eigencycle set 
       is defined to represent the set of $N(N-1)/2$ eigencycle elements. 
       The superscript $(mn)$ is the index of the two-dimensional subspace
       where the elements (eigencycles) of the set are located.
        $(mn)$ is defined as:
       \{$\{m,n\} \in \{1,2,...,n\} \cap (m < n)$\}.
       The assignment order is $m$ from 1 to $N$ first and then $n$ from 2 to $N$ in this paper.
\item \textbf{The parameter selection}
      Referring to the definition of the eigencycle, 
      changing $a$ in the game matrix $A$, 
      the values of the 6 eigencycles will be changing. 
      The results are shown in Fig. \ref{fig:ec_a}.   
      Following points is the reasons we choose the parameters. 
      (1) In order to be able to verified in real human subject experiments, 
      the parameter need to be simple and understandable.  
      (2) Various parameter means various treatments, and the theoretical expectations needs having significant different between
      the various treatments. 
      As the result, we choose the $a$ = [1/4, 4] as the two treatments, 
      to which we will focus to investigate.
\item \textbf{Geometric presentation:}
      Fig. \ref{fig:lissa} illustrate the ideal cyclic motion of the replicator dynamics (see Eq. (\ref{eq:repliequl})) for the game (see Table \ref{tab:gameRAmatrix}) with $a=[1/4, 4]$ near the Nash equilibrium. The geometric presentation of an eigencycle is similar
      to (1:1)-Lissajous diagrams.
      In (1:1)-Lissajous diagram, the amplitude
      of two components' amplitude  can be arbitrary,
      but the two components' amplitude of an eigencycle of a dynamics system at equilibrium 
      is fixed and not arbitrary due to the natural constrain of the eigenvectors.
      At the same time, the eigencycle only depends on the internal components
      $\eta_m$ and $\eta_n$ belonging to the unique complex eigenvector. 
      Referring to \cite{2019Roussel}, the cycle can be regarded as the projection of the eigen-trajectory in the 2 dimensional Euclidean spaces. 
\end{itemize}

The rotation axis can be a measurement for a 4-strategy game, 
because of a constrain condition of state space of one population game
is $\sum_{i=1}^4 x_i = 1$. So, the trajectory of the dynamics processes
can be fully presented by 3 dimensional variables ($x_1,x_2,x_3$). 
In this study, the rotation axis is the vector, 
which is defined as the same as the vector of angular momentum. 
That is the rotation axis vector components is defined as 
exactly equalling to 
the angular momentum vector components. 
A explanation of the definition of the measurement, as well as of the calculation approach for theoretical results are shown in Appendix   \ref{app:rot_measure}.  

Along the same procedure illustrated form Eq. (\ref{eq:repliequl}) to  Eq. (\ref{eq:sigmamatrix}), we can calculate the eigencycle values as well as the axis direction of a given dynamics equation system. Besides the replicator dynamics which labelled as [$T_1$], we select the MS-replicator dynamics (which labelled as [$T_2$], called also as adaptive replicator dynamics) and select logit dynamics (call also as noise best responses dynamics). In logit dynamics, to illustrate the dynamics pattern referring to the noise level, we select three noise parameter ([0.001, 0.05, 300]) which labelled as [$T_3, T_4, T_5$] respectively. 

In sum, for the five dynamics models (and parameters), the theoretical expectations can be calculated respectively. As the results, for the eigencycle set, the theoretical expectations are shown in Table \ref{tab:ec_result}  in the rows labelled as [$T_1, T_2, T_3, T_4, T_5 $].
And for the rotation axis vector components of the theoretical expectations are shown in Table \ref{tab:ax_result} in the rows labelled as [$T_1, T_2, T_3, T_4, T_5 $].

%
%

\subsection{Results from human subject experiment}\label{sec:res_human} 
We conducted the laboratory human subjects game experiment to investigate the game dynamics structure. There are two treatments of the experiment. The parameter of $a$ in games are 1/4 and 4 respectively. There are 8 sessions repeated for each treatments. Each session includes 1000 periods repeated game, which last about 2.5  hours to 3 hours. Average payment of each subject is 150 Yuan RMB. Each session has 6 human subject participated.  The game matching protocol is randomly match, in which every periods the counterpart of a subject is randomly  selected  from oneof the other 5 subjects, which is the same as \cite{wang2014social}. Details of the experimental protocol see the Appendix \ref{app:humanexp}. 

There is 8000 rounds time series for each of the two treatment. We use the time series to measure the eigencycles and the rotation axis direction  for each of the two treatment $a= [1/4, 4]$.  

\begin{figure}[!ht] 
\centering 
\includegraphics[scale=0.6]{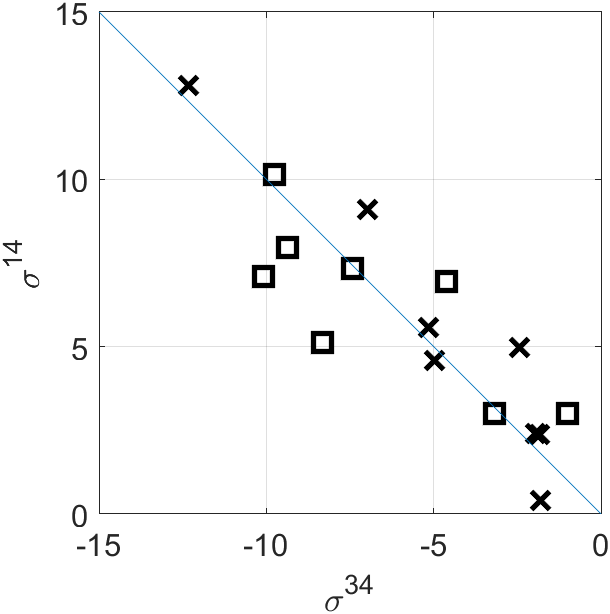} ~~~~~~~
\includegraphics[scale=0.6]{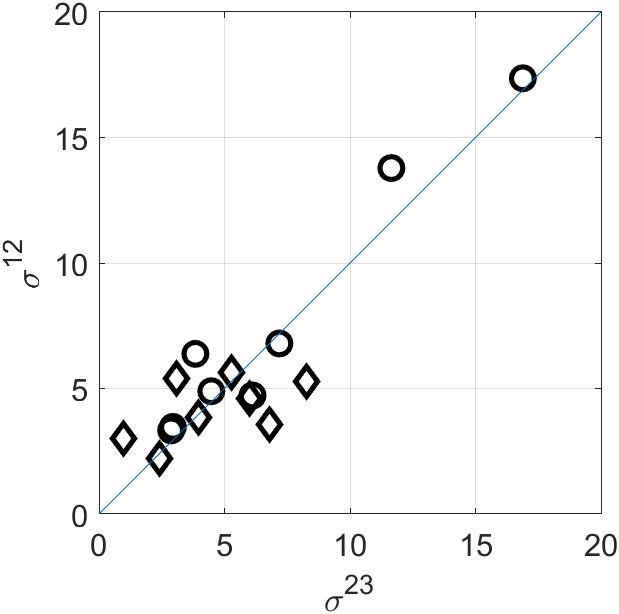}  
\caption{The presentation of relationship of the independence of $a$ between the  observations. Scatter experiment eigencycle values shown in left panel supports the relationship predicted in Eq. (\ref{eq:srtict_pm1}), in which the  square indicates $a=1/4$ treatment and  the  cross indicates $a=4$ treatment.  Scatter experiment eigencycle values shown in right panel supports the relationship predicted in Eq. (\ref{eq:srtict_pm2}), in which the cycle indicates $a=1/4$ treatment and  the  diamond indicates $a=4$ treatment. Notice that, as each treatment has 8 sessions repeated, each scattering has 8 samples.}
\label{fig:strict_pm}
\end{figure}

\begin{itemize}
\item For the eigencycle set, the results of the human subject game experiments are shown in Table \ref{tab:ec_result} in the row labelled as [$E$]. 
\item And for the rotation axis vector components, the results are shown in Table \ref{tab:ax_result} in the row labelled as [$E$].
\item As a responses to strict relationship of the independence of $a$ shown in Eq. (\ref{eq:srtict_pm1}) and in Eq. (\ref{eq:srtict_pm2}), we show the relationship 
from the data in Fig. \ref{fig:strict_pm}. Obviously, the relationships hold in significant.
\item  As a responses to strict relationship of the independence of $a$ shown in Eq. (\ref{eq:srtict_pm0}),  statistical results show that the prediction can not be rejected by data ($ttest$, $p$=0.6029, sample size $N$=8 in $a=1/4$ treatment; $ttest$, $p$=0.2239, sample size $N$=8 in $a=4$ treatment).  
\end{itemize}

\begin{table} 
\caption{The eigencycles of theory ($T$), human experiment ($E$) and simulation ($S$) \label{tab:ec_result}}
\begin{center}

\begin{tabular}{l|rrrrrrr}
     	 \hline 
     &	$\sigma_{12}$& $\sigma_{13} $& $\sigma_{14}$& $\sigma_{23}$&	$\sigma_{24}$& $\sigma_{34}$\\
     	 \hline 
$E$:  Human Exp. &	&	&	&	&	&	\\ 	
~~~~~~$a$=1/4&	0.0046&	0.0021&	-0.0067&	0.0042&	0.0004&	0.0063\\ 	
~~~~~~$a$=	4&	0.0070&	-0.0024&	-0.0047&	0.0076&	-0.0006&	0.0053\\ 
     	 \hline 						
$S_1$: Replicator&	&	&	&		&	&	\\ 
~~~~~~$a$=	1/4&	0.0004&	0.0002&	-0.0006&	0.0004&	0&	0.0006\\ 
~~~~~~$a$=	4&	0.0002&	-0.0001&	-0.0001&	0.0002&	0&	0.0001\\ 
$S_2$: MSReplicator&	&	&	&		&	&	\\ 
~~~~~~$a$=	1/4&	0.0565&	0.0073&	-0.0639&	0.0560&	0.0005&	0.0633\\ 
~~~~~~$a$=	4&	0.0728&	-0.0212&	-0.0516& 	0.0723&	0.0005&	0.0511\\ 
$S_3$: Logit[0.001]&	&	&	&		&	&	\\ 
~~~~~~$a$=	1/4&	0.0023&	0.0009&	-0.0032&	0.0026&	-0.0003&	0.0035\\ 
~~~~~~$a$=	4&	0.0091&	-0.0047&	-0.0044& 	0.0089&	0.0002&	0.0042\\ 
$S_4$: Logit[0.05]&	&	&	&		&	&	\\ 
~~~~~~$a$=	1/4&	0.0021&	0.0008&	-0.0030&	0.0025&	-0.0003&	0.0033\\ 
~~~~~~$a$=	4&	0.0081&	-0.0042&	-0.0039& 	0.0079&	0.0001&	0.0037\\ 
$S_5$: Logit[300]&	&	&	&		&	&	\\ 
~~~~~~$a$=	1/4&	0.1735&	-0.8947&	0.7212&	-0.7455&	0.919&	-1.6402\\ 
~~~~~~$a$=	4&	1.5904&	-0.4227&	-1.1677&	0.1526&	1.4378&	-0.2701\\ 
     	 \hline 						
$T_1$ Replicator &	&	&	&		&	&	\\ 
~~~~~~$a$=	1/4&	0.4659&	0.2795&	-0.7454& 	0.4659&	0&	0.7454\\ 
~~~~~~$a$=	4&	0.8653&	-0.5192&	-0.3461&	0.8653&	0&	0.3461\\ 
$T_2$ MSReplicator&	&	&	&		&	&	\\
~~~~~~$a$=	1/4&	0.4659&	0.2795&	-0.7454&	0.4659&	0&	0.7454\\ 
~~~~~~$a$=	4&	0.8653&	-0.5192&	-0.3461&	0.8653&	0&	0.3461\\ 
$T_3$ Logit[0.001]&	&	&	&		&	&	\\
~~~~~~$a$=	1/4&	0.4648&	0.2809&	-0.7457&	0.4658&	-0.0010&	0.7467\\ 
~~~~~~$a$=	4&	0.8662&	-0.5202&	-0.3461&	0.8658&	0.0004&	0.3457\\ 
$T_4$ Logit[0.05]&	&	&	&		&	&	\\
~~~~~~$a$=	1/4&	0.4133&	0.3160&	-0.7292&	0.4796&	-0.0663&	0.7957\\ 
~~~~~~$a$=	4&	0.9072&	-0.5527&	-0.3547&	0.8790&	0.0284&	0.3263\\ 
$T_5$ Logit[300]&	&	&	&		&	&	\\
~~~~~~$a$=	1/4&	0.5282&	0.1730&	-0.7012&	0.7576&	-0.2294&	0.9306\\ 
~~~~~~$a$=	4&	0.9455&	-0.3220&	-0.6234&	0.7331&	0.2123&	0.4111\\ 
     	 \hline						
\end{tabular}

\end{center} 
\end{table} 

\begin{table} 
\caption{The rotation axis vector components of theory ($T$), human experiment ($E$) and simulation ($S$) \label{tab:ax_result}}
\begin{center}
\begin{tabular}{l|rrr|rrr}
     	 \hline 
     ~~~~~~~~ Treatment	&	&$a$=1/4	&	&&	$a$=4	&	\\
~~~~~~~~~~~~~~~~Axis &	1~~~~&	3~~~~&	2~~~~&	1~~~~&	3~~~~&	2~~~~\\
     	 \hline 
Analytical & & &	 &&& \\
$T_1$: Replicator&	-0.0212&	-0.0212&	0.0127&	-0.0847&	-0.0847&	-0.0508\\
$T_2$: MSReplicator&	-0.1483&	-0.1483&	0.089&	-0.2754&	-0.2754&	-0.1653\\
$T_3$: Logit[0.001]&	-0.0001&	-0.0001&	0.0001&	-0.0004&	-0.0004&	-0.0002\\
$T_4$: Logit[0.05]&	-0.2038&	-0.1756&	0.1342&	-0.7953&	-0.8209&	-0.5\\
$T_5$: Logit[300]&	-0.0001&	-0.0001&	0&	-0.0001&	-0.0002&	-0.0001\\
     	 \hline 
 Simulation & & &	 &&& \\
$S_1$: Replicator &	-0.0004&	-0.0004&	0.0002&	-0.0002&	-0.0002&	-0.0001\\
$S_2$:  MSReplicator&	-0.056&	-0.0565&	0.0073&	-0.0723&	-0.0728&	-0.0212\\
$S_3$:  Logit[0.001]&	-0.0026&	-0.0023&	0.0009&	-0.0089&	-0.0091&	-0.0047\\
$S_4$:  Logit[0.05]&	-246.2&	-214.5&	81.87&	-790.9&	-805.9&	-417.8\\ 
$S_5$:  Logit[300]&	0.7455&	-0.1735&	-0.8947&	-0.1526&	-1.5904&	-0.4227\\
     	 \hline 
Human Exp. & & &	 &&& \\
$E$: mean &	-0.0042&	-0.0046&	0.0021&	-0.0076&	-0.007&	-0.0024\\
     	 \hline 
\end{tabular}
\end{center} 
\end{table}

\subsection{Results from agent-based simulation}\label{sec:res_simu} 
In order to investigate the dynamics structure, we hope having results from agent-based reinforcement learning models. In this study, we use the Agent-based evolutionary dynamics (ABED) simulator \cite{2019abed}, which is widely used in the field to study evolutionary game dynamics. 
  The platform has integrated various learning rules and matching rules, and has covering mainstream dynamics model of evolutionary dynamics, which is  an ideal platform to simulate the dynamics process for various models. 
  
  As mentioned above, there are 5 model (the replicator dynamics which labelled as $S_1$, MS-replicator dynamics  which labelled as $S_2$, and the three noise parameter [0.001, 0.05, 300] logit dynamics models  which labelled as $S_3, S_4, S_5$ respectively) for the two ($a-[1/4, 4]$) treatments. So we have 10  independent simulations protocol, respectively.  For each protocol, there is $10^5$ rounds time series. We use the time series to measure the eigencycles and the rotation axis direction.   Details of the protocols of the simulations are shown in Appendix \ref{app:agent}. 
  
Label the agent-based simulation with the protocol following the replicator dynamics setting  as ($S_1$) , the MS replicator dynamics setting as ($S_2$),  and the logit dynamics setting with noise parameter $[0.001, 0.05, 300]$ as as ($S_3, S_4, S_5$), from the time series, the results are reported.    
\begin{itemize}
\item For the eigencycle set, the results of the simulation are shown in Table \ref{tab:ec_result} in the rows labelled as [$S_1, S_2, S_3, S_4, S_5 $].
\item 
For the rotation axis vector components, the results are shown in Table \ref{tab:ax_result}  in the rows labelled as [$S_1, S_2, S_3, S_4, S_5 $].
\end{itemize}

\subsection{Consistency of theory and experiment}\label{sec:res_Consistency}
The consistency of the dynamics structure between the experiment, theory and simulation 
are central question of this study. To answer this question, we have identified the dynamics structure by eigencycle (result shown in \ref{tab:ec_result})
and the direction of the rotation axis (result shown in \ref{tab:ax_result}). 
Now, we calculates the correlation coefficients of the observations 
(the eigencycle set and the direction of the rotation axis) of the 
the experiment, theory and simulation
to report the statistical analysis results of the consistency.  

\begin{itemize}
\item On eigencycle measurement, the consistency between the theory and experiment 
is well in significant. The supporting data is following.  
\begin{itemize}
\item The eigencycle set of  the experiment, the 5 theory model and 5 agent based models simulation in the Table \ref{tab:ec_result}. 
We calculate the correlation coefficients for the two treatments $a=[1/4, 4]$ respectively. For $a=[1/4]$, the results are reported in Table \ref{tab:corr25}; and for  $a=[4]$, the results are reported in Table \ref{tab:corr44}. For visibility, we strike out the coefficients smaller than 0.900.
\item It is obvious that, except for extremely high noise condition $S_5$ and $T_5$ (noise parameter is 300) of the logit dynamics model, the experiment and  theory and simulation are consistence in significant  ($\rho > 0.900$, $N=6$). Importantly, the experiment results can be well interpreted by the models well in $a=[1/4]$ treatment (see the first column in Table \ref{tab:corr25})  
and in $a=[4]$ treatment (see the first column in Table \ref{tab:corr44}). 
The consistency of theory and experiment is supported in strongly significant ($\rho > 0.950$, $N=6$). 
\item Fig. \ref{fig:ec_te} illustrate the relationship of the normalised theoretical and the  normalised experimental eigencycles for the two ($a=[1/4, 4]$) treatments. 
By ordinary linear regression, the match between the theory and experiment is well in significant ($p=0.000$ for $a=[1/4]$ treatment, $p=0.003$ for $a=[4]$ treatment,  the sample size of each treatment $N = 6$).
\end{itemize}

\begin{figure}[!ht]
\centering
\includegraphics[width=4in]{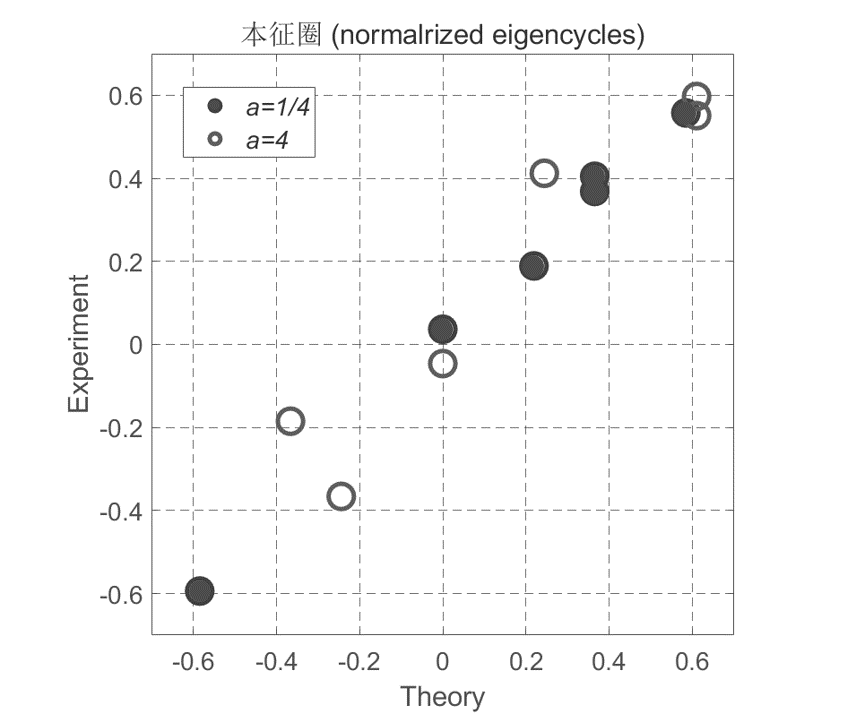}
\caption{Relationship of the normalised theoretical and the  normalised experimental eigencycles for the two ($a=[1/4, 4]$) treatments. The normalised theoretical eigencycles results comes from the replicator dynamics model $T_1$. The normalisation is that, the six components is divided by the root of the sum of their square. For $T_2,T_3,T_4$ and $S_1,S_2,S_3,S_4$, the relationship are similar. }
\label{fig:ec_te}
\end{figure}

\item On rotation direction axis vector components measurement, the march between the theory and experiment 
is well in significant. The supporting data is following. 
\begin{itemize}
\item The rotation direction axis vector components of  the experiment, the 5 theory model and 5 agent based models simulation in the Table \ref{tab:ax_result}.  
We calculate the correlation coefficients for the two treatments $a=[1/4, 4]$ respectively. For $a=[1/4]$, the results are reported in Table \ref{tab:corr25_ax}; and for  $a=[4]$, the results are reported in Table \ref{tab:corr44_ax}. For visibility, we strike out the coefficients smaller than 0.900.
\item It is obvious that, except for extremely high noise conditions $[S_5,T_5]$ (noise parameter is 300) of the logit dynamics model, all the theory and simulation are consistence well. Importantly, the experiment results can be well interpreted by the models well in $a=[1/4]$ treatment (see the first column in Table \ref{tab:corr25_ax})   
and in $a=[4]$ treatment (see the first column in Table \ref{tab:corr44_ax}) in Appendix section \ref{sec:add_sta}. 
\end{itemize} 
\end{itemize}
In sum, these two measurements provides same conclusion that, in general, 
in the human subject game experiments, the game dynamics structure
can be captured by the game dynamics models in significant.

\begin{table} 
\caption{The correlation  coefficients  of eigencycles of theory ($T$), human experiment ($E$) and simulation ($S$) for $a=[1/4]$ \label{tab:corr25}}
\begin{center}
\begin{tabular}{c|rrrrrrrrrrr}
     	 \hline   
& $E$ & $S_1$ &$S_2$&	$S_3$&	$S_4$&	$S_5$&	$T_1$ &$T_2$&	$T_3$&	$T_4$&	$T_5$\\
     	 \hline 
$E$&1&	&	&	&	&	&	&	&	&	&	\\
$S_1$&0.998&	1&	&	&	&	&	&	&	&	&	\\
$S_2$&0.980&	0.979&	1&	&	&	&	&	&	&	&	\\
$S_3$&0.990&	0.995&	0.986&	1&	&	&	&	&	&	&	\\
$S_4$&0.988&	0.993&	0.986&	1.000&	1&	&	&	&	&	&	\\
$S_5$&\sout{-0.697}&	\sout{-0.733}&	\sout{-0.640}&	\sout{-0.751}&	\sout{-0.755}&	1&	&   &	&	&	\\
$T_1$&0.997&	0.999&	0.969&	0.991&	0.990&	\sout{-0.748}&	1&	&	&	&	\\
$T_2$&0.997&	0.999&	0.969&	0.991&	0.990&	\sout{-0.748}&	1.000&	1&	&	&	\\
$T_3$&0.997&	0.999&	0.969&	0.991&	0.989&	\sout{-0.749}&	1.000&	1.000&	1&	&	\\
$T_4$&0.987&	0.993&	0.955&	0.989&	0.988&	\sout{-0.804}&	0.996&	0.996&	0.996&	1&	\\
$T_5$&0.953&	0.966&	0.967&	0.986&	0.988&	\sout{-0.801}&	0.960&	0.960&	0.960&	0.968&	1\\ 
     	 \hline

\end{tabular}
\end{center} 
\end{table} 

\begin{table} 
\caption{The correlation coefficients of eigencycles of theory ($T$), human experiment ($E$) and simulation ($S$) for $a=[4]$  \label{tab:corr44}}
\begin{center}
\begin{tabular}{c|rrrrrrrrrrr}
     	 \hline   
& $E$ & $S_1$ &$S_2$&	$S_3$&	$S_4$&	$S_5$&	$T_1$ &$T_2$&	$T_3$&	$T_4$&	$T_5$\\
     	 \hline 
$E$  & 1&	&	&	&	&	&	&	&	&	&	\\
$S_1$& 0.980&	1&	&	&	&	&	&	&	&	&	\\
$S_2$& 0.997&	0.976&	1&	&	&	&	&	&	&	&	\\
$S_3$& 0.973&	0.999&	0.971&	1&	&	&	&	&	&	&	\\
$S_4$& 0.973&	0.999&	0.970&	1.000&	1&	&	&	&	&	&	\\
$S_5$ & \sout{0.482}&	\sout{0.552}&	\sout{0.546}&	\sout{0.568}&	\sout{0.566}&	1&	&   &	&	&	\\
$T_1$ &0.953&	0.994&	{0.947}&	0.996&	0.996&	\sout{0.547}&	1&	&	&	&	\\
$T_2$&0.953&	0.994&	{0.947}&	0.996&	0.996&	\sout{0.547}&	1.000&	1&	&	&	\\
$T_3$&0.953&	0.994&	{0.947}&	0.996&	0.996&	\sout{0.548}&	1.000&	1.000&	1&	&\\
	$T_4$&{0.944}&	0.991&	{0.940}&	0.994&	0.995&	\sout{0.570}&	0.999&	0.999&	0.999&	1&	\\
	$T_5$ & 0.954&	0.968&	0.972&	0.969&	0.968&	\sout{0.720}&	{0.949}&	{0.949}&	{0.949}&	{0.950}&	1\\
     	 \hline

\end{tabular}
\end{center} 
\end{table}

\section{Discussion}\label{sec:diss}

This report illustrates that, the dynamics behaviours in human subject game experiment and game dynamics theory are consistence. The main contribution of this paper includes: (1) Confirmed that the non-Euclidean superplane cycle is real for the first time. (2) Suggest the finding from O'Neill game \cite{ONeill1987} and reported in \cite{WY2020} is not a coincidence; The eigencycle approach validates in 4-strategy game experiment case too. (3) Confirmed that the motion characteristics of the cycle are truly predictable, observable and controllable.

\begin{figure} 
\centering 
\includegraphics [width=6.0in]{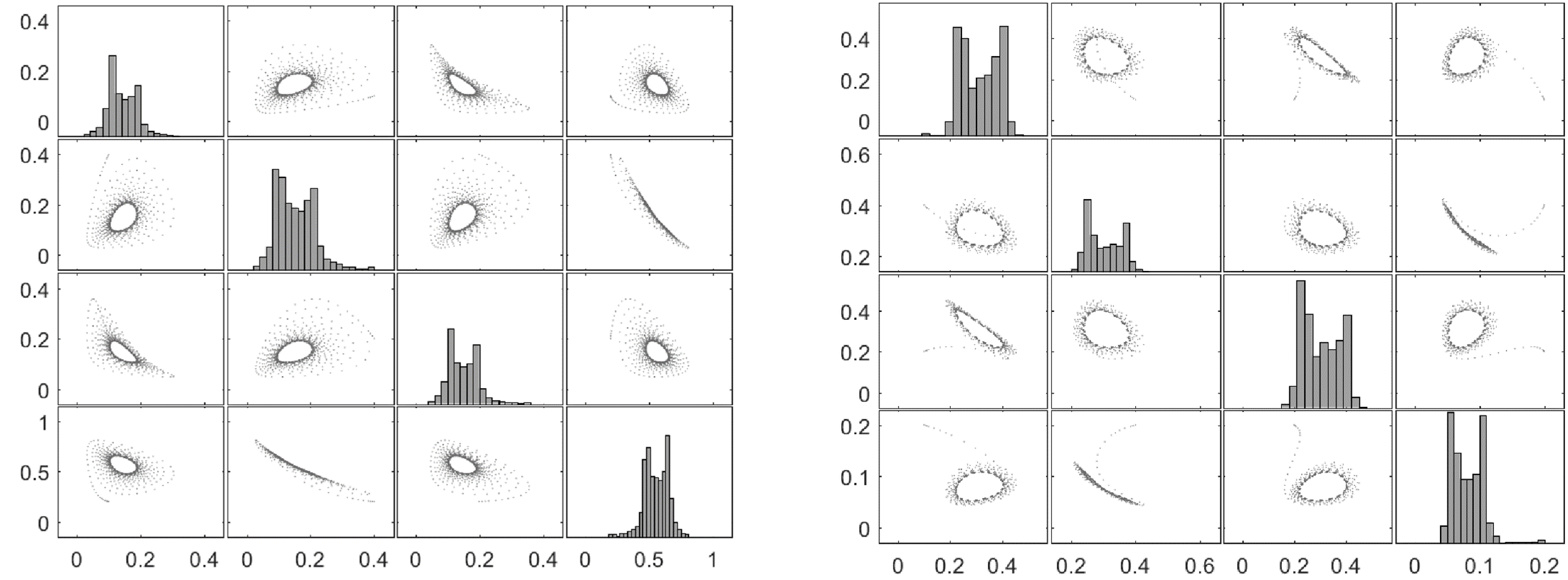}\\    
\caption{Matrix scatter plot of a sample evolutionary trajectory projected to the 2-dimensional subspace of the state space. The trajectory (time series) generated by the replicator dynamics equations with random initial condition. It is obviously that, so called as cycle, the closed period orbit is not in a Euclidean plane. In both treatments, in the ($x_2,x_4$)-subspece, the projections of the orbit are not straight line segments, but curve segments. Left and right panel are of treatment $a=[1/4,4]$ respectively. }
\label{fig:a4_manifold}
\end{figure}

\textbf{On related works} --- To our knowledge, till now, the match of the theory and experiment has not got rid of the constraints of Euclidean 2-dimensional space so far. Past 10 years has seen the quantitatively matching between theory and experiment on game dynamics, but of the all published works, the theoretical expectations are actuarial of Euclidean 2-d  \cite{dan2010tasp}\cite{wang2014}\cite{dan2014}\cite{wang2014social}. Even in the 4-strategy games experiments \cite{dan2010tasp,1999Huyck}, the concerned cycles in Euclidean plane. In the price dynamics cycle investigation, the price as strategy are continuous, but as results, the theoretical expectation and experimental measurement are projected to Euclidean 2-d plane to verify \cite{dan2021price}). 

The game selected in this study is of superplane (or twisted plane), not of ordinary 2 dimension Euclidean plane. This can be seen Fig \ref{fig:a4_manifold}, in which, the projection of the persistent cycle is not a straight line segment but a curve line segment. Both of $a=[1/4, 4]$ treatment have same performance. So, we call the cycle is of superplane cycle.  

\textbf{On further work} --- In the human subject experiment, we have notice that, there exists a structural difference between the experimental strategy distribution, which deviates from theoretical expectation show in Eq. (\ref{eq:nash_equilibrium}). This is a puzzle. Details of the distribution deviation observed in experiments can refer to \cite{2021Shujie} . And second, the axis of rotation or the direction of rotation under different parameters can only be qualitatively differentiated by using angular momentum. How to identify the segment's curve rate (which is the theoretical expectation shown in Fig. (\ref{fig:a4_manifold}) in experiment is unclear. These issues, all together, are relating to the consequence of the evolution trajectory on distribution of a game. Naturally, out of metaphor, more investigates in real strategy interactions systems would make game dynamics becoming more accurate, understandable and applicable.     
 
\textbf{On dynamics structure control} --- We realise the control of the dynamics structure by control the parameter $a$. This difference from the existed literature \cite{dan2010tasp}\cite{wang2014}\cite{dan2014}\cite{wang2014social}, in which, the object of the control is the eigenvalue for various stability. In this study, the object of the control is the eigenvector. We control the cycles structure by controlling the payoff matrix element $a$. To our knowledge, this report has provided a new realisation of game mechanism design for dynamics structure. Control-by-design is a critical issue not only in engineering, in game theory it is a field namely mechanism design. 

Considering the dynamics structure (eigenvector, invariant manifold, eigen mode) associates to business cycle in macroeconomics and microeconomics \cite{1987eigenvectors,2020relationship}, the control of dynamics structure is not trivial issue. During this manuscript publication process, we have notice that,  using the 5-strategy game \cite{WY2020} experiments as the benchmark, using pole assignment approach in modern game theory which is a state depended closed loop feedback design, the control of the dynamics structure is realisable in significant \cite{wang2022}. A noticeable point is that, in this research, when we change the control variable $a$ of the game, the dynamics structure is changed, but admittedly the equilibrium of game is changed too. Alternatively, in the closed loop feedback control \cite{wang2022}, when the dynamics structure changed, the equilibrium is not changed.  At the same time, no additional  finance is needed during controlled game dynamics processes. 

\section{Appendix}\label{sec:method}
 
Methods used in this study includes: (1) five game dynamic system equations, (2) agent-based simulation of evolutionary game dynamics and (3) laboratory human subjects game experiment. At the same time, in measurement we use  (1) the direction of the axis of rotation and (2) the eigencycle. In this section we will introduce these methods in details. 

\subsection{The five game dynamics models\label{app:dynEqs}}
There are five models with parameters applied to illustrate the match between theory and experiments. They are (1) the replicator dynamics which labelled as [$T_1$] in main text, (2) the MS-replicator dynamics which is labelled as [$T_2$]  in main text, and (3) logit dynamics (call also as noise best responses dynamics). In logit models, we select three noise parameter ([0.001, 0.05, 300]) which are labelled as [$T_3, T_4, T_5$]  in main text, respectively.  
The dynamics equations are presented as following. 
 
\begin{itemize}
\item 
Replicator Dynamics 
\begin{equation}
    \dot{x}_i=x_i (U_i -\overline{U} ), 
\end{equation} 
in which $\dot{x}_i$ is the velocity of the proportion growth of the population using $i$-th strategy, $U_i$ is the payoff of an agent in the population using $i$-th strategy; $\overline{U}$ indicates the mean payoff of the population. 
 In main text, this model denoted as $[T_1]$.
For one population symmetric game which payoff matrix being A, we have  
\begin{equation}\label{eq:u_i}
    U_i = \sum_{j=1}^N A_{ij} x_j 
\end{equation}  
and 
\begin{equation}\label{eq:u_bar}
    \overline{U} = \sum_{i=1}^N x_i U_i
\end{equation}  
\item 
MS Replicator Dynamics is an adjusted replicator dynamics.  
\begin{equation}\label{eq:msrepli_dyn}
     \dot{x}_i=\frac{x_i (U_i -\overline{U})}{\overline{U}}           
\end{equation}  
In which $\dot{x}_i$ is the velocity of the population using $i$, $U_i$ is the payoff of an agent in the population $i$; $\overline{U}$ indicates the mean payoff in the population.
The algorithm for $U_i$ and $\overline{U}$ see Eq. (\ref{eq:u_i}) and Eq. (\ref{eq:u_bar}). In main text, this model denoted as $[T_2]$.
\item 
Logit dynamics is noise best response model. 
\begin{equation}\label{eq:logit_dyn}
             \dot{x}_i= \frac{\exp(\lambda U_i)}{\sum_{j=1}^N \exp(\lambda U_j)} - x_i  
\end{equation}  
In which  $\lambda$ is the noise parameter; $\dot{x}_i$ is the velocity of the population using $i$, $U_i$ is the payoff of an agent in the population $i$.  The algorithm for $U_i$ see Eq. (\ref{eq:u_i}). 
 In main text, the $[T_3,T_4,T_5]$ relates to $\lambda = [0.001, 0.05, 300]$ condition respectively. 
\end{itemize}

\subsection{Human subject game experiment protocol\label{app:humanexp}}

The experiment was approved by the Experimental Social Science Laboratory of Zhejiang University. The data of controlled treatment ($a=[1/4,~4]$-Treatment) is from the experiment carried out from Nov. to Dec 2020.  The authors confirms that this experiment was performed in accordance with the approved social experiments guidelines and regulations, which follow the regulation of \textbf{experimental economics} protocol \cite{Behavioral2003}. 

A total number of 96 undergraduate and graduate students of Zhejiang University volunteered to serve as the human subjects of this experiment. These students were openly recruited through a web registration system. 

The 96 human subjects (call also as players) were distributed into 16 populations of equal size $N$ = 6. The six players of each population carried one experimental sessions (see session organisation Table). During the game process the players sited separately in a classroom laboratory, each of which facing a computer screen. They were not allowed to communicate with each other during the whole experimental session. Written instructions were handed out to each player and the rules of the experiment were also orally explained by an experimental instructor. The rules of the experimental session are as follows:
\begin{enumerate}
\item 
Each player plays the 4-strategy game repeatedly with the same other five players.

\item 
Each player earns virtual points during the experimental session according to the payoff matrix shown in the written instruction.  

\item 
In each game round, each player competes with one player in the other five players as the opponent.

\item 
Each player has  to make a choice among the four candidate actions “x1”, “x2”, “x3” , “x4” . If this time runs out, the player has to make a choice immediately. After a choice has been made it can not be changed.
\end{enumerate}
 
\begin{figure}
\centering
\includegraphics[width=0.5\textwidth]{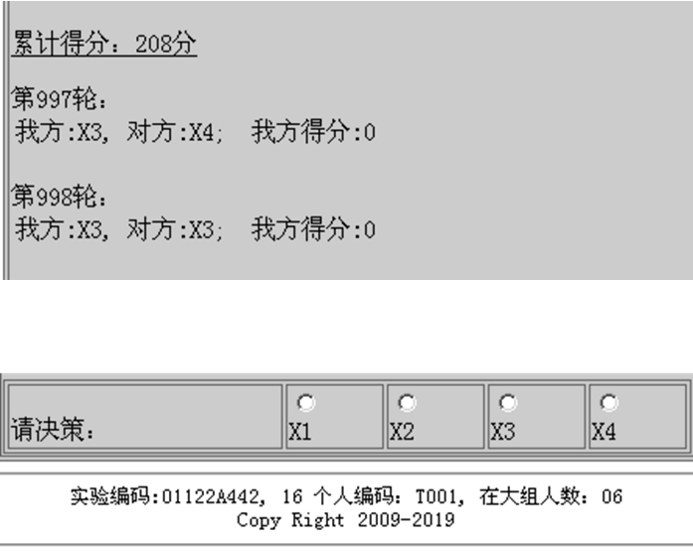}
\caption{\label{fig:expInterface} The screen shot of the user interface in human subject game experiment. }
\end{figure}

\begin{table} 
\caption{The experiment session organisation}
\begin{center}
\begin{tabular}{|cccc|cccc|}
 \hline 
$a$=1/4&	&	&		&	$a$=4&	&	&	\\
 SessionID&	Date& Subjects&Period&			SessionID&	Date&	Subjects&	Period\\
 \hline  
01121A251&	20201121&	6&	1000&		01122A441&	20201122&	6&	1000\\
01121A252&	20201121&	6&	1000&		01122A442&	20201122&	6&	1000\\
01122A253&	20201122&	6&	1000&		01128A443&	20201128&	6&	1000\\
01122A254&	20201122&	6&	1000&		01128A444&	20201128&	6&	1000\\
01128A255&	20201128&	6&	1000&		01129A445&	20201129&	6&	1000\\
01129A256&	20201129&	6&	1000&		01129A446&	20201129&	6&	1000\\
01219A258&	20201219&	6&	1000&		01219A448&	20201219&	6&	1000\\
01227A250&	20201227&	6&	1000&		01228A449&	20201228&	6&	1000\\

 \hline
\end{tabular}
\end{center}
\label{tab:ezp}
\end{table}

During the experimental session, the computer screen of each player will show an information window and a decision window. The window on the left of the computer screen is the information window. The upper panel of this information window shows the current game round, the time limit (40 seconds in controlled  ($a = [1/4, 4]$-Treatment)) of making a choice and the time left to make a choice. The color of this upper panel turns to green at the start of each game round.  After all the players  have made their decisions, the lower panel of the information window will show the player's own choice, the opponent strategy, and own payoff in this game round are shown in the screen. The player's own accumulated payoff is also shown. The players are asked to record their choices of each round on the record sheet in some round for checking. Each session last 2.5-3hours, have more 1000 periods records from a session. For each $a=[1/4, 4]$ treatment, we have 8 sessions repeated. So, we have totally 8000 records in the time series for each treatment.   

The window on the right of the computer screen is the decision window. It is activated only after all the players of the group have made their choices. The upper panel of this decision window lists the current game round, while the lower panel lists the four candidate actions (“x1”, “x2”, “x3” , “x4”) horizontally from left to right. The player can make a choice by clicking on the corresponding action names.  

The reward for each player is determined by the rank, which is determined by the total number of their earning points in experiment sessions participated.  Form the highest to the lowest, each player is payed as 275, 225, 175, 125, 75 and 25 yuan RMB in controlled treatments.

\subsection{Agent-based evolutionary dynamics simulation protocol\label{app:agent}}
 Reinforcement learning theory is a branch of game theory. Agent-based evolutionary dynamics simulation is an approach to understand the consequence of reinforcement learning theory. Method of computer simulation to evaluate the consistency of theory and experiment is introduce as following. 
\begin{description}
  \item[1. Select simulation platform:] We use abed simulator \cite{2019abed}, which is widely used in the field to study evolutionary game dynamics. 
  The platform has integrated various learning rules and matching rules, and has covering mainstream dynamics model of evolutionary dynamics, which is  an ideal platform to simulate the dynamics process. The platform is of the long-running repeated game setting 
  in finite populations.  
    \item[2. Setting parameters] The parameter setting of the five simulation are shown in Table \ref{tab:sim_para}. The authors has carefully classify the (approximate) 
    equivalent 
    between dynamics evolutionary equations and the parameter setting for simulation. For example, for replicator dynamics model, the simulation is under imitative protocols, in which  candidates are agents; meanwhile, the decision method is of pairwise-comparison of the strategy payoff. The complete-matching is set. These setting are follow the user guide of the platform, which system will performs like replicator dynamics shown in Eq. (\ref{eq:repliequl}) in large population (1000 agents) and low reversion probability (1\%) limit.
  \item[3. Conduct the simulation:]  
      In our study case, for each of the 5 models and each of the treatments investigated, we need to run a 1M period simulation. The time cost for each run of the simulation of a given parameter set is about 30 minutes in a desktop personal computer, which CPU is 8  GHz and the memory is 16 GB. 
  \item[4. Analysis the time series:]  Main outcome of the simulator is the time series. The time series, including the strategies density and their payoffs,  can be outputted from the platform in detail. These can be used to evaluate the performance of the controller-by-design, for example fluctuation, as well as the efficiency, profits or social welfare evolution along time. 
\end{description}

\begin{table} 
\caption{The parameter setting for the 5 models simulations}\label{tab:sim_para}
\begin{center}
\begin{tabular}{|r|rrrrrr|}
     	 \hline
Parameter&	Replicator $[S_1]$ &	& MS-Replicator $[S_2]$&	&	Logit $[S_3, S_4, S_5]$&	\\
     	 \hline
payoff-matrix&	[[ 0 0 0 4 ]~ &	& as left &	&	as left &	\\
             &	 [ 1 0 0 0 ]~ &	&   &	&	  &	\\
             &   [ 0 1 0 0 ]~ &	&   &	&	  &	\\
             &	 [ 0 0 1 0 ]] &	&   &	&	  &	\\
n-of-agents&	1000&	&	1000&	&	1000&	\\
random-initial-condition?&	FALSE &	&	FALSE &	&	FALSE &	\\
initial-condition&	[250 250 250 250] &	&	[250 250 250 250] &	&	[250 250 250 250] &	\\
candidate-selection&	imitative  &	&	imitative  &	&	imitative  &	\\
n-of-candidates&	2&	&	2&	&	2&	\\
decision-method&	pairwise- &	&	positive- &	&	logit&	\\
 &	difference &	& proportional &	&	&	\\
complete-matching?&	TRUE &	&	TRUE &	&	TRUE &	\\
n-of-trials&	999&	&	999&	&	999&	\\
single-sample?&	TRUE  &	&	TRUE  &	&	TRUE  &	\\
tie-breaker&	uniform &	&	uniform &	&	uniform &	\\
log-noise-level&	0&	&	0&	&	 0.001~~$S_3$ &	\\
 &	0&	&	0&	&	 0.05~~~$S_4$&	\\
 &	0&	&	0&	&	 300~~~~$S_5$ &	\\
use-prob-revision?&	TRUE &	&	TRUE &	&	TRUE &	\\
prob-revision&	0.2&	&	0.2&	&	0.2&	\\
n-of-revisions-per-tick&	500&	&	500&	&	500&	\\
prob-mutation&	0.002&	&	0.002&	&	0.002&	\\
trials-with- replacement?&	FALSE &	&	FALSE &	&	FALSE &	\\
self-matching?&	FALSE &	&	FALSE &	&	FALSE &	\\
imitatees-with- replacement?&	FALSE &	&	FALSE &	&	FALSE &	\\
consider-imitating-self?&	FALSE  &	&	FALSE  &	&	FALSE  &	\\
plot-every-?-secs&	2&	&	2&	&	2&	\\
duration-of-recent&	10&	&	10&	&	10&	\\
show-recent-history?&	TRUE &	&	TRUE &	&	TRUE &	\\
show-complete-history?&	TRUE&	&	TRUE&	&	TRUE&	\\
     	 \hline 
\end{tabular}
\end{center} 
\end{table}

\subsection{Angular momentum as the measurement}

According to the theoretical the eigencycle set decomposition approach,
we can carry out the cyclic angular momentum measurement
in  each of the two-dimensional subspace,
indicated by the eigencycle $\Omega^{(mn)}$, separately.
%
The angular momentum $L^{(mn)}_E$\cite{wang2017} can be expressed by the following formula: \\
\begin{equation}\label{eq:exp_am}
    L^{mn}_E=\frac{1}{N-1}\sum_{t=1}^{N-1}\left(x(t)-O\right) \times \left(x(t\!+\!1)-x(t)\right)
\end{equation}

\begin{itemize}
\item $L^{(mn)}_E$ represents the average value of
the accumulated angular momentum over time;
the subscript $mn$ indexes the two-dimensional $(x_m,x_n)$ subspace;
\item $N$ is the length of the experimental time series, that is,
the total number of  repetitions of the repeated game experiments;
\item $O$ is the projection of the Nash equilibrium at the subspace $\Omega^{(m,n)}$;
\item $x(t)$ is a two-dimensional vector at time $t$
which can be expressed as $(x_m(t),x_n(t))$, and $x(t+1)$ is at time $t+1$;
\item $\times$ represents the cross product between two two-dimensional vectors.
\end{itemize}
This measurement can be called also as signed area of
the triangle $\Delta_{[O, x(t), x(t+1)]}$ in the ($m,n$) 2-d subspace.
For each transition from $x(t)$ to $x(t+1)$ referring to $O$,
the angular momentum is twice of the signed area of the triangle.
We suggest using the concept of the angular momentum,
because it contain
the mass $m$ as parameter,
which may compatible the population size $N$
as variable in further game dynamics investigation. 

\subsection{Rotation axis as the measurement}\label{app:rot_measure}
The axis of rotation is the direction of the 3 dimensional angular momentum. Considering $\sum_{i=1}^4 x_i =1$, we ignore the $x_4$ and remain $(x_1,x_2,x_3)$ as independent variable. So, it turns out to be a 3-d issue. 
Angular momentum is the area swept by a vector in unit time. 
In two dimensional motion case, 
the direction is perpendicular to the 2 dimensional plane. 
In order to better observe the direction of the axis in the three-dimensional space, the 1st, 3rd, and 2nd dimension components of the eigenvectors are selected for calculation, and the obtained angular momentum is a three-dimensional vector. Selecting 1, 3, and 2 components for theoretical calculation in order corresponds to simulation data analysis and graphical analysis. It is planned to set strategy 1 as the x-axis, strategy 3 as the y-axis, and strategy 2 as the z-axis. The formula for theoretical calculation of angular momentum is as follows \cite{2021Shujie}:
\begin{equation}
    \overline{L} = \frac{1}{T} \int_0^T \Re{(x(t))} \times \Re{(v(t))} 
\end{equation} 
Here, $x(t)$ represents of strategy vector $x$ at time $t$, $v(t)$ represents the instantaneous speed of observation $x(t)$, $\Re$ means takeing the real part. $\times$ means the cross-multiplication. $\overline{L}$ is the  mean angular momentum between time $[0, T]$.  
The theory analytical results of the axis of rotation uses the eigenvector components to calculate the angular momentum. For each model,  
an arbitrary small (for example $10^{-5}$) deviates from the Nash equilibrium, and select $T\longrightarrow \infty$. It is not difficult to prove that, this measurement equivalent to the angular momentum measurement in Eq. (\ref{eq:exp_am}) which can be applied in the measurements with simulation and experiment time series.


\subsection{Additional statistic test \label{sec:add_sta}}
This part is a supplementary information of statistical results for the main text.  
 
\begin{itemize}
    \item Table \ref{tab:corr25_ax} is the correlation  coefficients  of the rotation axis  for $a=[1/4]$; 
    \item Table \ref{tab:corr44_ax} is the correlation  coefficients  of the rotation axis  for $a=[4]$.
\end{itemize}

\begin{table} 
\caption{The correlation  coefficients  of the rotation axis  for $a=[1/4]$ \label{tab:corr25_ax}}
\begin{center}
\begin{tabular}{c|rrrrrrrrrrr}
          	 \hline   
&$T_1$ &$T_2$&	$T_3$&	$T_4$&	$T_5$ & $S_1$ &$S_2$&	$S_3$&	$S_4$&	$S_5$&	 $E$\\
     	 \hline 
$T_1$&    	 1 & & & & & & & & & & \\
$T_2$&       1.000 & 1 & & & & & & & & & \\
$T_3$&       1.000 & 1.000 & 1 &  & & & & & & & \\
$T_4$&       0.997 & 0.997 & 0.997 & 1 & & & & & & & \\
$T_5$&       1.000 & 1.000 & 1.000 & 0.997 & 1 & & & & & & \\
$S_1$&       1.000 & 1.000 & 1.000 & 0.997 & 1.000 & 1 & & & & & \\
$S_2$&       1.000 & 1.000 & 1.000 & 0.997 & 1.000 & 1.000 & 1 & & & & \\
$S_3$&       	 0.997 & 0.997 & 0.997 & 1.000 & 0.997 & 0.997 & 0.996 & 1 & & & \\
$S_4$&       	 0.996 & 0.996 & 0.996 & 1.000 & 0.996 & 0.996 & 0.996 & 1.000 & 1 & & \\
$S_5$&       	 \sout{-0.829} & \sout{-0.829} & \sout{-0.829} & \sout{-0.869} & \sout{-0.829} & \sout{-0.829} & \sout{-0.825} & \sout{-0.870} & \sout{-0.875} & 1 & \\
$E$&       	 0.999 & 0.999 & 0.999 & 0.992 & 0.999 & 0.999 & 0.999 & 0.991 & 0.990 & \sout{-0.798} & 1 \\
     	 \hline 
\end{tabular}
\end{center} 
\end{table} 

\begin{table} 
\caption{The correlation  coefficients  of the rotation axis for $a=[4]$ \label{tab:corr44_ax}}
\begin{center}
\begin{tabular}{c|rrrrrrrrrrr}
     	 \hline   
&$T_1$ &$T_2$&	$T_3$&	$T_4$&	$T_5$ & $S_1$ &$S_2$&	$S_3$&	$S_4$&	$S_5$&	 $E$\\
     	 \hline 
$T_1$&     	 1 & & & & & & & & & & \\
$T_2$&     	 1.000 & 1 & & & & & & & & & \\
$T_3$&     	 1.000 & 1.000 & 1 & & & & & & & & \\
$T_4$&     	 0.997 & 0.997 & 0.997 & 1 & & & & & & & \\
$T_5$&     	 \sout{0.500} & \sout{0.500} & \sout{0.500} & \sout{0.561} & 1 & & & & & & \\
$S_1$&     	 1.000 & 1.000 & 1.000 & 0.997 & \sout{0.500} & 1 & & & & & \\
$S_2$&     	 1.000 & 1.000 & 1.000 & 0.998 & \sout{0.507} & 1.000 & 1 &  & & & \\
$S_3$&     	 0.999 & 0.999 & 0.999 & 1.000 & \sout{0.534} & 0.999 & 0.999 & 1 & & & \\
$S_4$&     	 0.999 & 0.999 & 0.999 & 0.999 & \sout{0.529} & 0.999 & 1.000 & 1.000 & 1 & & \\
$S_5$&     	 \sout{0.339} & \sout{0.339} & \sout{0.339} & \sout{0.406} & 0.984 & \sout{0.339} & \sout{0.347} & \sout{0.377} &\sout{ 0.371} & 1 & \\
$E$&     	 0.994 & 0.994 & 0.994 & 0.984 & \sout{0.406} & 0.994 & 0.993 & 0.989 & 0.990 & \sout{0.238} & 1 \\
     	 \hline 
\end{tabular}
\end{center} 
\end{table}




\end{document}